\documentclass[12pt]{article}
\begin{document}
\title{Quantum Decoherence and Pointer Basis:\\ Dynamics in State Vectors}%

\author{{Kentaro Urasaki}%
\footnote{E-mail: urasaki@f6.dion.ne.jp}\\
Kamiigusa 3-5-15-103, Suginami, Tokyo, Japan}\date{}%

\maketitle

{\abstract
It is well-known that the pointer basis of a quantum system 
satisfies the condition to diagonalize the interaction Hamiltonian between the subsystems.
We show that this condition can be translated into the form $\delta\Lambda=0,$ 
where $\Lambda$, so-called the action, is the time integrated interaction energy: 
it is found out naturally in the phase of state vectors due to diagonal interaction. 
The careful treatment of a two states system demonstrates that the states of the 
total system branch into the states with different values of the action. 
Mathematically the pointer states are selected out 
by the saddle point condition on the phase $\Lambda$.
This study helps us to understand the precise mechanism and the general dynamics of decoherence.}

\section{Introduction}
The concept of decoherence is expected to reproduce not only the framework of classical physics but also classicality observed in our experience, including the arrow of time. 
The recent developement of this field has clarified the important role 
of the environment for quantum systems. 
Although it is still difficult task to establish a new view of nature, our understanding about `openness' has certainly advanced for the last three decades by many contributions on this subject. 

So far, however, the important consequences of decoherence have been understood and discussed mainly with the use of density matricies (for example, \cite{Joos1985, Zurek1981, Zurek1982}). 
Needless to say, it is desirable that these consequences can be interpreted also in terms of state vectors that represent the established kinematical concept of quantum theory\cite{Zeh1997}.
In this study we intend to reconstruct the argument that leads decoherence and in particular pointer basis: the set of privileged states to express a system interacting with its environment\cite{Zurek1981}. 
Consequently, we show that the main consequences can be derived also from the state vectors.

\section{Extension of state space due to action}
We start from the initial state in the simple product form of normalized vectors, $|\Phi(t_0)\rangle=|\phi(t_0)\rangle|\varepsilon(t_0)\rangle$, where 
$|\phi(t_0)\rangle$ denotes the state of the apparatus entangled with the microscopic system and $|\varepsilon(t_0)\rangle$ denotes its environment. 
These non-perturbative states are defined to obey Schr\"odinger equations: 
\begin{eqnarray}
&&[i\hbar\partial_t-\hat{h}_\phi]|\phi(t)\rangle=0, \label{eq:subsystem1}\\
&&[i\hbar\partial_t-\hat{h}_\varepsilon]|\varepsilon(t)\rangle=0,\label{eq:subsystem2}
\end{eqnarray}
where $\hat{h}_\phi$ ($\hat{h}_\varepsilon$) is the Hamiltonian of the subsystem 
to act only on $|\phi(t)\rangle$ ($|\varepsilon(t)\rangle$).
Then, the non-perturbative state of the total system $|\Phi_0(t)\rangle=|\phi(t)\rangle|\varepsilon(0)\rangle$ also satisfies, 
\begin{eqnarray}\label{eq:whole0}
[i\hbar\partial_t-\hat{h}_\phi-\hat{h}_\varepsilon]|\Phi_0(t)\rangle=0. 
\end{eqnarray}

Let us consider the effect of the interaction between these systems, $\hat{h}_{int}$, in the Schr\"odinger equation for the total system: 
\begin{eqnarray}\label{eq:whole}
[i\hbar\partial_t-\hat{h}_\phi-\hat{h}_\varepsilon-\hat{h}_{int}]|\Phi(t)\rangle=0. 
\end{eqnarray}
First, we do the calculation simply assuming that we can use the basis of the non-perturbative system, $\{|\Phi_{0n}(t)\rangle\}$, which has completeness at each time $t$, to expand a state of the total system even when the interaction, $\hat{h}_{int}$, exists:
this means that we can expand a perturbed state the same way as the standard perturbation theory,
\begin{eqnarray}\label{eq:expansion}
|\Phi(t)\rangle=\sum_nC_n(t)|\Phi_{0n}(t)\rangle.
\end{eqnarray}
Substituting this into Eq. (\ref{eq:whole}) and acting $\langle\Phi_{0n}(t)|$, 
we obtain 
\begin{eqnarray}
i\hbar\partial_tC_n(t)=\sum_{n^\prime}C_{n^\prime}(t)\langle\Phi_{0n}(t)|\hat{h}_{int}|\Phi_{0n^\prime}(t)\rangle.
\end{eqnarray}
In many cases one may replace $C_{n^\prime}(t)$'s in the right hand side with $C_{n^\prime}(t_0)$'s under the assumption of the weak time dependence for these coefficients. 

In order to consider the interaction between the macroscopic systems, 
we, however, must interpret that $\{|\Phi_{0n}(t)\rangle\}$ represents the basis consisting of macroscopically distinguishable states.
Then the following two points are crucially important:\\
(1) Even if the interaction is too weak to occur the macroscopic transition between 
these states, the time integration of the interaction energy 
over a finit interval gives cosiderably large contribution compared with $\hbar.$\\
(2) The number of these states is much smaller than that the total system potentially has. \\
Let us analyze these facts in detail below: 
we see the validity of the lowest order (mean-field) approximation 
for the variation of the state vectors, in which we can exactly treat the time dependence originating from the interaction energy.

(1) Because the interaction is weak not to occur the macroscopic transition, 
it is reasonable that $\hat{h}_{int}$ is considered to be approximately diagonal in a certain basis $\{|\Phi_{0n}(t)\rangle\}$ as Zurek did\cite{Zurek1981}: 
$\langle\Phi_{0n}(t)|\hat{h}_{int}|\Phi_{0n^\prime}(t)\rangle\simeq\delta_{n,n^\prime}\langle\Phi_{0n}(t)|\hat{h}_{int}|\Phi_{0n}(t)\rangle.$
Therefore we obtain,
\begin{eqnarray}
i\hbar\partial_tC_n(t)\simeq C_n(t)\langle\Phi_{0n}(t)|\hat{h}_{int}|\Phi_{0n}(t)\rangle\label{eq:diagonal}.
\end{eqnarray}
We can easily integrate it as $C_n(t)=C_n(t_0)e^{-i\Lambda_n(t)/\hbar}$, where $\Lambda_n=\int^t_{t_0}\langle\Phi_{0n}(t)|\hat{h}_{int}|\Phi_{0n}(t)\rangle dt$ and, 
\begin{eqnarray}\label{eq:expansion2}
|\Phi(t)\rangle\simeq\sum_nC_n(t_0)|\Phi_{0n}(t)\rangle e^{-i\Lambda_n(t)/\hbar}
\end{eqnarray}
Therefore if the initial state is in the eigenstate, $|\Phi(t_0)\rangle=|\Phi_{0n}(t_0)\rangle,$ this state evolves into $|\Phi_n(t)\rangle=|\Phi_{0n}(t)\rangle e^{-i\Lambda(t)_n/\hbar}$. 
The standard perturbation theory teaches us 
that $C_{n^\prime}(t)$'s in the right hand side 
can be replaced by $C_{n^\prime}(t_0)$'s for weak interaction in many cases.
The factor $\Lambda_n/\hbar$, however, becomes so large 
for the contact of the macroscopic systems in the present case that 
we cannot neglect the time dependence of $C_{n^\prime}(t)$'s.

We here notice that the solution above obtained, $|\Phi_n(t)\rangle=|\Phi_{0n}(t)\rangle e^{-i\Lambda_n(t)/\hbar}$ satisfies the mean-field equation:
\begin{eqnarray}
[i\hbar\partial_t-\hat{h}_\phi-\hat{h}_\varepsilon-\langle\Phi_n(t)|\hat{h}_{int}|\Phi_n(t)\rangle]|\Phi_n(t)\rangle=0.
\end{eqnarray}
As easily understood, if the non-perturbative state $|\Phi_{0n}(t)\rangle$ is also the exact eigenstate of the interaction Hamiltonian, $\hat{h}_{int}$, at time $t$, the mean-field solution $|\Phi_n(t)\rangle$ agrees with the exact solution of the original equation, Eq. (\ref{eq:whole}) at this moment. 
(For example, a localized state is the eigenstate of the Coulomb interaction but it instantly spread over obeying Eq. (\ref{eq:whole}).)
And only in this case, the assumption Eq. (\ref{eq:expansion}) is correct in the strict sense. 
In general, it is obvious that the basis of the system satisfies Eq. (\ref{eq:diagonal}) holds only in the sense of the mean-field approximation although this approximation can be justifiable in the case that for a macroscopic system interacting with its environment. 

(2) Therefore it is a consistent approach to adopt the mean-field approximation for an arbitrary non-perturbed state: the state evolves into, 
$|\Phi(t)\rangle=|\Phi_0(t)\rangle e^{-i\Lambda_\Phi(t)/\hbar}$, which obviously satisfies 
\begin{eqnarray}\label{eq:mean-field}
[i\hbar\partial_t-\hat{h}_\phi-\hat{h}_\varepsilon-\langle\Phi(t)|\hat{h}_{int}|\Phi(t)\rangle]|\Phi(t)\rangle=0, 
\end{eqnarray}
where $\Lambda_\Phi=\int^t_{t_0} \langle\Phi(t)|\hat{h}_{int}|\Phi(t)\rangle dt.$
From the fact that this equation has the non-linearlity depending on $|\Phi(t)\rangle$, 
we necessarily reconsider the expansion of the total system with the basis $\{|\Phi_n(t)\rangle\}.$
The interaction $\hat{h}_{int}$ requires us to extent the state space.

We then should consider carefully the superposition of the product states paying attention to both the linearity of the equation (\ref{eq:whole}) and the time dependence of $\Lambda(t)$.
Two non-orthogonal initial states, $\langle\Phi_0(t_0)|\Phi_0(t_0)^\prime\rangle\ne 0,$ evolve into the states, $|\Phi(t)\rangle=|\Phi_0(t)\rangle e^{-i\Lambda_\Phi(t)/\hbar}$ and $|\Phi^\prime(t)\rangle=|\Phi_0^\prime(t)\rangle e^{-i\Lambda_{\Phi^\prime(t)}/\hbar}$ with different $\Lambda$'s in general.
Therefore these must be treated as the {\bf linearly independent solutions} of Eq. (\ref{eq:whole}) because the time dependence of $\Lambda_\Phi-\Lambda_{\Phi^\prime}$ leads the orthogonality relation, 
\begin{eqnarray}
\int^\infty_{-\infty}\langle\Phi(t)|\Phi^\prime(t)\rangle dt=\langle\Phi_0(t)|\Phi_0^\prime(t)\rangle\int^\infty_{-\infty}e^{i(\Lambda_\Phi-\Lambda_{\Phi^\prime})/\hbar}dt=0. 
\end{eqnarray}
(We used the fact that the coefficient $\langle\Phi_0(t)|\Phi_0^\prime(t)\rangle$ is independent of time as immediately represented in the eigenstates of the non-perturbed Hamiltonian, $\{|\Phi_{0\epsilon}\rangle\}$. The expansions, $|\Phi_0(t)\rangle=\sum_\epsilon C_\epsilon|\Phi_{0\epsilon}\rangle e^{-i\epsilon t/\hbar}$ and $|\Phi_0^\prime(t)\rangle=\sum_\epsilon C^\prime_\epsilon|\Phi_{0\epsilon}\rangle e^{-i\epsilon t/\hbar}$ lead $\langle\Phi_0(t)|\Phi_0^\prime(t)\rangle=\sum_\epsilon C^\ast_\epsilon C_\epsilon^\prime$.)
We here assumed that $\Lambda$'s monotonously increase because 
we consider the continuous interaction between the apparatus and its environment. 
It is already known that this time dependent `action', $\Lambda$, plays important role when decoherence occurs\cite{Zurek1981, Zurek1982}. It, however, has been discussed mainly in the information context.

\section{Two states system}
In the previous section we found that the lowest-order (particular) solution is constructed with the solutions of (\ref{eq:subsystem1}) and (\ref{eq:subsystem2}) in the product form as, 
\begin{eqnarray}\label{eq:action}
|\Phi(t)\rangle=|\Phi_0(t)\rangle e^{-i\Lambda(t)/\hbar}=|\phi(t)\rangle|\varepsilon(t)\rangle e^{-i\Lambda(t)/\hbar},
\end{eqnarray}
and here $\Lambda(t)$ is the action: 
\begin{eqnarray}
\Lambda(t)=\int^t_{t_0}\langle \phi(t)|\hat{V}|\phi(t)\rangle dt
=\int^t_{t_0}\langle \Phi(t)|\hat{h}_{int}|\Phi(t)\rangle dt,
\end{eqnarray}
where $\hat{V}=\langle \varepsilon(t)|\hat{h}_{int}|\varepsilon(t)\rangle$ is the `external' field that continuously effects on $|\phi\rangle$.
We also found that the orthogonality relation in terms of the time integral emerges from non-orthogonal initial states. 

Let us study further the case that the apparatus, represented by $|\phi(t_0)\rangle$, is the two states system. Firstly we consider the following linear combination in order to find out one particular solution of Eq. (\ref{eq:whole}): 
\begin{eqnarray}
|\phi_\theta(t_0)\rangle=\cos\theta|\phi_\uparrow(t_0)\rangle+\sin\theta|\phi_\downarrow(t_0)\rangle,
\end{eqnarray}
where $|\phi_\uparrow(t_0)\rangle$ and $|\phi_\downarrow(t_0)\rangle$ are two normalized eigenstates of $\hat{h}_\phi$ and orthogonal each other.
That is to say, the state of the apparatus is represented by the superposition of two macroscopically distinct states at $t_0$.
For an arbitrary $\theta,$ each pair $|\phi_\theta(t_0)\rangle$ and $|\phi_{\theta+\pi/2}(t_0)\rangle$ can be the basis of this subsystem. 

We here assume the diagonal form of the interaction especially for the basis $|\phi_0(t_0)\rangle=|\phi_\uparrow(t_0)\rangle$ and $|\phi_{\pi/2}(t_0)\rangle=|\phi_\downarrow(t_0)\rangle$, so that the Hamiltonian of the subsystem and the interaction Hamiltonian commute, $[\hat{h}_\phi, \hat{V}]=0$ (so-called non-demolition case\cite{Zurek1981}).
For simplicity, although originally such a diagonal form for a macroscopic system is an approximate relation as mentioned in the previous section, we do calculation if it was an exact relation here: we assume 
\begin{eqnarray}\label{eq:off-diagonal}
\langle \phi_\uparrow(t_0)|\hat{V}|\phi_\downarrow(t_0)\rangle=\langle  \phi_\downarrow(t_0)|\hat{V}|\phi_\uparrow(t_0)\rangle=0.
\end{eqnarray}
Moreover we also assume the interaction energy is independent of the state of the environment, $|\varepsilon(t)\rangle$. 
These lead,
\begin{eqnarray}
\Lambda_\theta(t)&=&\cos^2\theta\int^t_{t_0}\langle  \phi_\uparrow(t)|\hat{V}|\phi_\uparrow(t)\rangle dt+\sin^2\theta\int^t_{t_0}\langle  \phi_\downarrow(t)|\hat{V}|\phi_\downarrow(t)\rangle dt\\
&\equiv&\cos^2\theta\Lambda_\uparrow(t)+\sin^2\theta\Lambda_\downarrow(t).
\end{eqnarray}
Therefore now the initial states simply evolves as, 
\begin{eqnarray}
|\Phi_\theta(t_0)\rangle=|\phi_\theta(t_0)\rangle|\varepsilon(t_0)\rangle \to |\Phi_\theta(t)\rangle=|\phi_\theta(t)\rangle|\varepsilon(t)\rangle e^{-i\Lambda_\theta(t)/\hbar}, 
\end{eqnarray}
where we can interpret that the environment only changes its phase from $|\varepsilon(t_0)\rangle$ to $|\varepsilon_\theta(t)\rangle=|\varepsilon(t)\rangle e^{-i\Lambda_\theta(t)/\hbar}$.

We here notice again that $|\Phi_\theta(t)\rangle$ depends on the coefficients $\cos^2\theta$ and $\sin^2\theta$ through the action. 
Therefore the linearity of the subsystem $|\phi\rangle$ cannot be extended for the total system $|\Phi(t)\rangle$ in its entirely:
For the initial state $|\phi(t_0)\rangle=(\alpha|\phi_0(t_0)\rangle+\beta|\phi_{\pi/2}(t_0)\rangle)|\varepsilon(t_0)\rangle,$ the time evolution in the form, $|\phi(t)\rangle=\alpha|\phi_\uparrow(t)\rangle|\varepsilon_\uparrow(t)\rangle+\beta|\phi_\downarrow(t)\rangle|\varepsilon_\downarrow(t)\rangle$, appears to be certainly correct as considered in the earlier studies\cite{Zurek1981, Zurek1982} (see also \cite{Scully1978}). 
There, however, is ambiguity in this expression:
When the interaction exists, states of the total system should be distinguished also by the time evolution through the action, $\Lambda$, as below.

Obviously $\Lambda_\uparrow(t)\ne\Lambda_\downarrow(t)$ reproduces 
the orthogonality relation $\langle \varepsilon_\theta(t)|\varepsilon_{\theta^\prime}(t)\rangle=e^{i(\Lambda_\theta(t)-\Lambda_{\theta^\prime}(t))/\hbar}\to\delta_{\theta,\theta^\prime}$ for {\bf time average}, which is consistent with the discussion by Zurek\cite{Zurek1982}. 
At the same time, we also find that the corresponding states of the total system, 
$|\Phi_\theta(t)\rangle$ and $|\Phi_{\theta^\prime}(t)\rangle$, are the 
linearly independent solutions of Eq. (\ref{eq:whole}) 
for $\Lambda_\theta(t)\ne\Lambda_\theta^\prime(t)$: 
\begin{eqnarray}\label{eq:orthogonality}
\int^\infty_{t_0}\langle \Phi_\theta(t)|\Phi_{\theta^\prime}(t)\rangle dt=\cos(\theta-\theta^\prime)\int^\infty_{t_0}dt e^{-i(\Lambda_\theta(t)-\Lambda_{\theta^\prime}(t))/\hbar}=\delta_{\theta,\theta^\prime}.
\end{eqnarray}
For example, $\displaystyle\frac{1}{\sqrt2}(|\phi_\uparrow(t)\rangle+|\phi_\downarrow(t)\rangle)=|\phi_{\pi/4}(t)\rangle$ and $|\phi_\uparrow(t)\rangle=|\phi_0(t)\rangle$ give different time dependence to the action, $\Lambda_\theta(t),$ and resultingly the corresponding total states $|\Phi_{\pi/4}(t)\rangle$ and $|\Phi_0(t)\rangle$ are {\bf orthogonal in time} each other.
If we assume the constant monitoring of macroscopic object by its environment, i.e., $\Lambda(t)=\lambda (t-t_0)$, we estimate that the orthogonality is achieved in very short time, $\tau\simeq \hbar/\lambda$, where the interaction energy, $\lambda=\langle\phi(t)|\hat{V}|\phi(t)\rangle$, is a macroscopic quantity. 

Therefore the interaction between subsystems fractionates the expansion of the total system through $\Lambda,$ so that the linear combination of $|\Phi_\theta(t)\rangle$'s, is {\bf necessarily} required. 
An arbitrary states at $t_0$ is expressed as $|\phi(t_0)\rangle=e^{i\alpha}\cos\theta|\phi_\uparrow(t_0)\rangle+e^{i\beta}\sin\theta|\phi_\downarrow(t_0)\rangle$, but $\Lambda$ is independent of these phase factors, $\alpha$ and $\beta$, in the present model.
Then, in the mean-field level, the (general) solution of Eq. (\ref{eq:whole}) is represented appropriately as, 
\begin{eqnarray}\label{eq:Phit}
|\Phi(t)\rangle&&=\sum_{0\le\theta\le\pi/2} C_\theta|\Phi_\theta(t)\rangle\\
&&=\sum_{0\le\theta\le\pi/2} C_\theta |\phi_\theta(t)\rangle|\varepsilon(t)\rangle e^{-i\Lambda_\theta(t)/\hbar}\\
&&=\sum_{0\le\theta\le\pi/2} C_\theta(\cos\theta|\phi_\uparrow(t)\rangle+\sin\theta|\phi_\downarrow(t)\rangle)|\varepsilon_\theta(t)\rangle,
\end{eqnarray} 
where $\sum_\theta|C_\theta|^2=1$. 
(Although the angle $\theta$ is continuous, here we use the notation $\sum_\theta$ just for simplicity. 
Explicitly we should use $\displaystyle\frac{2}{\pi}\int^{\pi/2}_0\cdots d\theta$ 
and the orthogonality relation is also changed.)

As a matter of course, also for the explicitly expressed initial state $|\Phi(t_0)\rangle=|\Phi_0(t_0)\rangle=\sum_\theta C_\theta|\Phi_\theta(t_0)\rangle,$ the total system evolves unitarily as,
\begin{eqnarray}\label{eq:unitarily}
e^{-i\hat{H}(t-t_0)/\hbar}|\Phi(t_0)\rangle&&=\sum_\theta C_\theta e^{-i\hat{H}(t-t_0)/\hbar}|\phi_\theta(t_0)\rangle|\varepsilon(t_0)\rangle\\
&&\simeq\sum_\theta C_\theta |\phi_\theta(t)\rangle|\varepsilon(t)\rangle e^{-i\Lambda_\theta(t)/\hbar},
\end{eqnarray}
under the mean-field approximation. 
We notice that only giving the initial conditions $\alpha$ and $\beta$ for the subsytem $|\phi(t)\rangle=\alpha|\phi_\uparrow(t)\rangle+\beta|\phi_\downarrow(t)\rangle$ is generally insufficient to determine the evolution of the whole system $|\Phi(t)\rangle$ with no ambiguity 
because the initial state, $|\Phi(t_0)\rangle=|\Phi_0(t_0)\rangle=\left[(\sum_\theta C_\theta \cos\theta)|\phi_\uparrow(t_0)\rangle+(\sum_\theta C_\theta \sin\theta)|\phi_\downarrow(t_0)\rangle\right]|\varepsilon(t)\rangle$, has a kind of degeneracy for $\theta$.
(The analysis in \cite{Zurek1982} starts from the particular initial states corresponding to single $\theta$.)
In other words, in appearance, different time evolutions can emerge from an identical initial condition, $|\phi(t)\rangle=\alpha|\phi_\uparrow(t)\rangle+\beta|\phi_\downarrow(t)\rangle$.

\section{decoherence in the state vectors: pointer basis}
For $t>t_0,$ the interaction between the macroscopic systems, $|\phi(t)\rangle$ and $|\varepsilon(t)\rangle,$ naturally makes the action $\Lambda(t)$ large. 
Under the condition $\Lambda_\theta(t)\gg\hbar$, we can adopt the saddle point approximation on the solution: the contribution of the terms that satisfy $\displaystyle \frac{\delta}{\delta\theta}\Lambda_\theta(t)=0$ only survive in equation (\ref{eq:Phit}),
\begin{eqnarray}
|\Phi(t)\rangle&\sim& \tilde{C}_0|\phi_\uparrow(t)\rangle|\varepsilon(t)\rangle e^{-i\Lambda_\uparrow(t)/\hbar}+
\tilde{C}_{\pi/2}|\phi_\downarrow(t)\rangle|\varepsilon(t)\rangle e^{-i\Lambda_\downarrow(t)/\hbar}\\
&=&\tilde{C}_0|\phi_\uparrow(t)\rangle|\varepsilon_\uparrow(t)\rangle+
\tilde{C}_{\pi/2}|\phi_\downarrow(t)\rangle|\varepsilon_\downarrow(t)\rangle,\label{eq:approximatevectors}
\end{eqnarray}
where $\tilde{C}_{0, \pi/2}=C_{0, \pi/2}\sqrt{\frac{2\pi i}{\Lambda_{0, \pi/2}^{\prime\prime}(t)/\hbar}}=C_{0, \pi/2}\sqrt{\frac{\pm\pi i\hbar}{\Lambda_\uparrow(t)-\Lambda_\downarrow(t)}}$.

Now we can conclude that the condition $\Lambda_\theta(t)\gg\hbar$ (more explicitly the variation of $\Lambda_\theta(t)/\hbar$ being large and also assumed $\Lambda_\uparrow(t)\ne\Lambda_\downarrow(t)$) 
leads that only one basis, $\{|\Phi_\theta(t)\rangle=|\phi_\theta(t)\rangle|\varepsilon_\theta(t)\rangle\}$, is privileged to express the total system $|\Phi(t)\rangle$:
This basis consists of the $|\Phi_\theta(t)\rangle$'s, which give extremum to the action, $\Lambda_\theta(t)$. The interaction Hamiltonian is naturally diagonal in this basis because the saddle point condition $\delta\Lambda=0$ always selects out the approximate eigen states of the interaction Hamiltonian from the all of the solutions $\{|\Phi_\theta(t)\rangle\}$.
We have demonstrated the emergence of pointer basis in terms of state vectors. 

The density matrix of the total system should be composed from the expression (\ref{eq:approximatevectors}) rather than (\ref{eq:Phit}). 
In the present context, the reduced density matrix of the subsystem $\rho_{SA}={\rm Tr}_\varepsilon|\Phi(t)\rangle\langle\Phi(t)|$ has the off-diagonal terms with the factor $r(t)=\langle\varepsilon_\uparrow(t)|\varepsilon_\downarrow(t)\rangle=e^{i(\Lambda_\uparrow(t)-\Lambda_\downarrow(t))/\hbar}.$ 
These will vanish for the time average\cite{Zurek1982}.

\section{Conclusion Remarks}
We saw the precise mechanism of decoherence in two states system in terms of state vectors. 
It is crucially important to recognize that the interaction between macroscopic systems requires the extension of the state space. 

The consequences in this study reproduce the discussion in terms of density matrices by Zurek\cite{Zurek1981,Zurek1982}. 
A kind of action principle leads a pointer basis: the saddle point condition, $\displaystyle\frac{\delta \Lambda_\theta}{\delta\theta}=0$, selects out the states being the closest to the eigenstates of the interaction Hamiltonian from the all of the solutions of the Schr\"odinger equation for the total system. 
On the other hand, other states disappear because of the destructive intereference due to the action, $\Lambda$. 

Although, in \S 2, we assumed the diagonal form of the interaction energy in Eq. (\ref{eq:mean-field}). 
This assumption can best be satisfied for the pointer basis and may be too rough for the other states. 
This is the future task.

This action brings not only the dynamics but also the non-unitary `appearance' in the time evolution of the state of the total system as, 
\begin{eqnarray}
|\Phi(t_0)\rangle=\left[(\sum_\theta C_\theta \cos\theta)|\phi_\uparrow\rangle+(\sum_\theta C_\theta \sin\theta)|\phi_\downarrow\rangle\right]|\varepsilon\rangle\\ \to
|\Phi(t)\rangle\sim \tilde{C}_0|\phi_\uparrow\rangle|\varepsilon\rangle e^{-i\Lambda_\uparrow/\hbar}+
\tilde{C}_{\pi/2}|\phi_\downarrow\rangle|\varepsilon\rangle e^{-i\Lambda_\downarrow/\hbar}
, 
\end{eqnarray}
although it still evolves in unitary fashion (see Eq. (\ref{eq:unitarily})). 

It is also important to unify the dynamics demonstrated in the two states systems here.
For this purpose, our simple interpretation using an action is worthwhile
because it naturally reflects the time evolutional feature of a system.

\section*{Acknowledgments}
The author would like to thank S. Sato for helpful comments on the manuscript.

\end{document}